\documentclass[twocolumn,pre]{revtex4}
\usepackage{amssymb}
\usepackage{amsfonts}
\usepackage{amsmath}
\usepackage{graphicx}

\begin{document}

\title{Minimal epidemic model considering external infected injection and
governmental quarantine policies: Application to COVID-19 pandemic}
\author{L. A. Rodr\'{\i}guez Palomino}
\email{rodrigl@cab.cnea.gov.ar}
\author{Adri\'{a}n A. Budini}
\email{adrianbudini@gmail.com}
\affiliation{Consejo Nacional de Investigaciones Cient\'{\i}ficas y T\'{e}%
cnicas(CONICET), Centro At\'{o}mico Bariloche, Avenida E. Bustillo Km 9.5,
(8400) Bariloche, Argentina, and Universidad Tecnol\'{o}gica Nacional
(UTN-FRBA), Fanny Newbery 111, (8400) Bariloche, Argentina.}
\date{\today }

\begin{abstract}
Due to modern transportation networks (airplanes, cruise ships, etc.) an
epidemic in a given country or city may be triggered by the arrival of
external infected agents. Posterior government quarantine policies are
usually taken in order to control the epidemic growth. We formulate a
minimal epidemic evolution model that takes into account these components.
The previous and posterior evolutions to the quarantine policy are modeled
in a separate way by considering different complexities parameters in each
stage. Application of this model to COVID-19 data in different countries is
implemented. Estimations of the infected peak time-occurrence and epidemic
saturation values as well as possible post-quarantine scenarios are analyzed
over the basis of the model, reported data, and the fraction of the
population that adopts the quarantine policy.
\end{abstract}

\maketitle

\section{Introduction}

The COVID-19 pandemic started in the Chinese state Hubei at the end of 2019,
and since then it propagated to almost all countries. Contrarily to an
epidemic developing in a local closed region, the external propagation was
triggered by the arrival of infected travelers that return from the other
countries. In fact, at a continental level, after Asia, the epidemic
propagated to Europe (with focus on Italy and Spain) and later to America
(USA and Latino American countries) and Africa.

After the initial propagation, different administrations adopted quarantine
policies. The main goal is to slow down the velocity of infected people
increase, which in turn should avoid the collapse of the national health
systems. This issue is of central importance because the disease mortality
is decreased under appropriate medical attention.

At the present time, for most of the countries, the epidemic did not reach
its maximal value. From a mathematical modeling perspective, the main
challenge is to estimate the posterior epidemic evolution taking into
account the quarantine policies and the available data. Magnitudes of
special interest are the maximal (accumulative) number of infected agents as
well as the epidemic peak and its time of occurrence. A qualitative
description of possible post-quarantine scenarios is also of interest. In
fact, while strict quarantine policies are effective for controlling the
epidemic propagation, its sustainability and economical cost are aspects
under discussion.

The mathematical modeling of epidemic propagation and evolution is a well
established topic \cite{murray,networks}. One of the simplest descriptions
is given by the so called Susceptible-Infectives-Recovered (SIR) models. The
three variables of interest (S, I and R) count the number of persons in each
possible state. A non-linear contribution (the product of S and I) governs
the transition from the healthy to the infected group. There exists a vast
and growing literature dealing with the COVID-19 pandemic description, which
is based on SIR like models and related ones (see for example
(non-exhaustive) Refs.~\cite%
{berlin,niteroi,Schaback,Ferguson,Zhang,Read,wang,ritter,ramos,yang,lopez,peng,garcia}%
). In contrast to standard SIR models, where saturation of an epidemic
occurs due to a dominant number of recovered agents, here saturation and
epidemic decay are mainly produced (in a first stage) by the quarantine
policies.

The goal of this contribution is to formulate a simple and minimal epidemic
mathematical model that takes into account both the external injection of
infected agents as well as the quarantine government policies. In contrast
to other approaches, we consider that the initial free epidemical growth and
its posterior constrained evolution (due to quarantine) must be described in
intrinsic separate ways. In fact, both stages are determined by different
mechanisms whose underlying complexities may be quite dissimilar. The
proposed evolution is defined with a minimal number of parameters whose
estimation follows from data fitting and also public information. In the
pre-quarantine period the dynamic is linear while in the post-quarantine
stage the model can be related to a linearized SIR model. The approach is
applied to available data of different European countries and Argentina. The
maximal number of infected agents as well as the size of the epidemic peak
are estimated in each case. A qualitative description of post-quarantine
scenarios in terms of the fraction of isolated population is also proposed.

The manuscript is outlined as follows. In Sec. II we present the model and
its derivation. In Sec. III it is applied in the description of COVID-19
data for different countries: Italy, Spain, France, United Kingdom, and
Argentina. In Sec. IV post-quarantine scenarios are discussed. The
Conclusions are provided in Sec. V. A mapping with SIR models is presented
in the Appendix.

\section{Epidemic model}

Due to the connectivity between diverse parts of the world (airplanes,
cruise ships, etc.), in most countries the beginning of the epidemic was
produced by travelers returning from an infected region. In this \textit{%
first period}, where no local governmental policy has been taken yet, the
growth of the (total) infected population $n(t)$ results from external
injection plus local contagions. These features depend on a complex way on
flight connectivity and local passenger displacements, which in turn may
lead to very different average behaviors. For example, the infected
population may grow in an exponential (Latino American countries) or
sub-exponential way (some European countries). Given the increase of
infected people, in a\textit{\ second period} local governments take
different policies such as total or partial quarantine (avoiding local
spread), jointly with the cancellation (partial or full) of infected
external injection. In our approach, due to their intrinsic different
origins, these two stages are modeled in a separate way.

\subsection{Modelling free grow stage under external infected injection and
local contagions}

At the beginning, the epidemic increase is produced by an external injection
of infected agents, which in turn may produce a local spreading of the
disease. At this stage, we propose the evolution%
\begin{equation}
\frac{dn(t)}{dt}=a(t)+r(t)n(t),\ \ \ \ \ 0\leq t\leq t_{q}.
\label{ModelShort}
\end{equation}%
The function $a(t)>0$ measures the velocity of external infected injection
(i.e. airplanes). Furthermore, the function $r(t)>0$ gives the rate of local
(regional) contagions. This evolution is taken between the initial time $%
t=0, $ the day before the arrival of the first infected agent, up to the
time $t_{q}$ where a government policy is taken. Time is measured in day. We
analyze different possibilities for the functions $a(t)$ and $r(t).$

\subsubsection{Pure imported cases}

One assumption is to consider that the local spreading of the disease gives
a small or null contribution. Thus, we consider the functions%
\begin{equation}
a(t)=at^{\nu -1}\exp \Big{(}c\frac{t^{\nu }}{\nu }\Big{)},\ \ \ \ \ \ \ \ \
r(t)=0.  \label{Model1}
\end{equation}%
Here, the rate constants $a$ and $c$ have units $1/(day)^{\nu },$ while $\nu
>0$ is a dimensionless positive parameter. This functional form of $a(t)$ is
justified from the solution it gives rise. Eq. (\ref{ModelShort}) can be
integrated as $n(t)=n(0)+\int_{0}^{t}dt^{\prime }a(t^{\prime }),$ with $%
n(0)=0.$ Therefore,%
\begin{equation}
n(t)=\frac{a}{c}\Big{[}\exp \Big{(}c\ \frac{t^{\nu }}{\nu }\Big{)}-1\Big{]}%
,\ \ \ \ \ 0\leq t\leq t_{q}.  \label{Sol1}
\end{equation}

\subsubsection{Imported cases plus local propagation}

Alternatively, we may take a nonvanishing contribution due to local
propagation,$\ r(t)\neq 0.$ We assume%
\begin{equation}
a(t)=at^{\nu -1},\ \ \ \ \ \ \ \ \ r(t)=ct^{\nu -1},  \label{Model2}
\end{equation}%
where as before the rate constants $a$ and $c$ have units $1/(day)^{\nu },$
while $\nu >0$ is again a dimensionless positive parameter. With these
definitions, the solution of Eq.~(\ref{ModelShort}) reads $[n(0)=0]$%
\begin{equation}
n(t)=\frac{a}{c}\Big{[}\exp \Big{(}c\ \frac{t^{\nu }}{\nu }\Big{)}-1\Big{]}%
,\ \ \ \ \ 0\leq t\leq t_{q}.  \label{Sol2}
\end{equation}

\subsubsection{A compromise solution}

We notice that the models (\ref{Model1}) and (\ref{Model2}) lead to exactly
the same solutions, Eqs. (\ref{Sol1}) and (\ref{Sol2}). Hence, their
underlying different mechanisms cannot be discerned from $n(t).$ On the
other hand, the solutions present the desired time behavior. In fact, in a
short time regime $n(t)\simeq (a/\nu )t^{\nu },$ in an intermediate regime, $%
n(t)\simeq (a/\nu )t^{\nu }[1+(1/2)(c/\nu )t^{\nu }+\cdots ],$ while in the
posterior long time regime $n(t)\simeq (a/c)\exp (ct^{\nu }/\nu ).$ The
parameter $\nu $ allows us to cover very different growing behaviors. A
standard exponential growth corresponds to $\nu =1.$

One may argue that the equality of the previous solutions only applies to
the special chosen functions $a(t)$ and $r(t)$ [Eqs.~(\ref{Model1}) and (\ref%
{Model2})]. This is certainly true. Nevertheless, in real experimental data
it is hard to separate between imported and local contributions. Thus, while
it is possible to chose other functional forms of $a(t)$ and $r(t),$ from
the data is very hard to distinguish between them. Thus, we model the
epidemic growth as%
\begin{equation}
n(t)\simeq \frac{\tilde{a}}{\tilde{c}}\Big{[}\exp \Big{(}\tilde{c}\ \frac{%
t^{\nu }}{\nu }\Big{)}-1\Big{]},\ \ \ \ \ 0\leq t\leq t_{q}.  \label{Sol}
\end{equation}%
This function corresponds to Eq.~(\ref{Sol1}) or Eq.~(\ref{Sol2}). As a
solution of compromise, the value of the parameters $\tilde{a},$ $\tilde{c},$
and $\nu $ are determined from the experimental data. These parameters take
in an effective way both the external injection as well as the local
propagation of the disease. In a roughly way, the parameter $\nu $\ can be
read as a measure of the flight and local spread networks complexities. The
time $t_{q}$ is taken from public information corresponding to each country.
Given that any change in social behavior does not occur instantaneously,
this time has an intrinsic and unavoidable uncertainty.

\subsection{Modelling evolution after quarantine government policies}

Given the global network information structure, in diverse countries (or
cities) the local governments dictated quarantine policies in order to
change the free epidemic growth [Eq.~(\ref{Sol})] and try to get saturation
in the total number of reported cases at the minimum number of cases
possible. In order to take into account this underlying change in the
epidemic dynamics, we split the total number of infected as%
\begin{equation}
n(t)=n_{I}(t)+n_{R}(t).  \label{Conservation}
\end{equation}%
The term $n_{R}(t)$ gives the number of infected agents that recovered their
health or died. Thus,\ $n_{I}(t)$ is the number of infected agents able to
propagate the disease. For times $t\leq t_{q},$ for the data we analyze in
the next section, it is possible to approximate%
\begin{equation}
n_{I}(t)\simeq n(t),\ \ \ \ \ \ n_{R}(t)\simeq 0,\ \ \ \ \ \ 0\leq t\leq
t_{q}.  \label{Aprox}
\end{equation}%
where $n(t)$ is given by Eq. (\ref{Sol}).

After the quarantine, we assume the evolution%
\begin{equation}
\frac{dn_{I}(t)}{dt}=r_{q}(t)n_{I}(t)-\frac{1}{\tilde{\tau}}n_{I}(t),\ \ \ \
\ t_{q}\leq t.  \label{eneI}
\end{equation}%
Here, the time $\tilde{\tau}$ is a parameter to determine. It can be read as
the average number of days that an infected agent is able to propagate the
disease in the quarantine period. The rate of local contagions $r_{q}(t)>0$
is proportional to the number of susceptible population able to be infected
in the quarantine period. We notice that the structure of Eq. (\ref{eneI})
[with an exponential $r_{q}(t)]$ can be derived from a linearization of a
non-linear SIR like model [see Appendix]. For $n_{R}(t)$ we take the
evolution%
\begin{equation}
\frac{dn_{R}(t)}{dt}=\frac{1}{\tilde{\tau}}n_{I}(t),\ \ \ \ \ t_{q}\leq t,
\label{eneERE}
\end{equation}%
which guarantees that in a long time limit all the population falls in this
group. From Eqs. (\ref{eneI}) and (\ref{eneERE}), and the relation (\ref%
{Conservation}), the total number of infected people evolves as%
\begin{equation}
\frac{dn(t)}{dt}=r_{q}(t)[n(t)-n_{R}(t)],\ \ \ \ \ t_{q}\leq t.
\label{ModelCorregido}
\end{equation}

Given that quarantine leads to saturation of $n(t),$ the function $r_{q}(t)$
must fulfill%
\begin{equation}
\lim_{t\rightarrow \infty }r_{q}(t)=0.  \label{C1}
\end{equation}%
In addition, at time $t=t_{q}$ we demand the continuity of the derivative of 
$n(t),$ which leads to%
\begin{equation}
\left. \frac{dn(t)}{dt}\right\vert _{t=t_{q}}=r_{q}(t_{q})n(t_{q}),
\label{C2}
\end{equation}%
where the left and right terms follow from Eqs. (\ref{Sol}) and (\ref%
{ModelCorregido}) respectively.

\subsubsection*{A saturation model}

The post-quarantine dynamics is completely defined after knowing the rate $%
r_{q}(t).$ In order to fulfill condition~(\ref{C1}) we propose the function%
\begin{equation}
r_{q}(t)=\lambda t^{\mu -1}\exp \Big{(}-\tilde{\gamma}\frac{t^{\mu }}{\mu }%
\Big{)},  \label{EreQ}
\end{equation}%
where $\tilde{\gamma}$ has units $1/(day)^{\mu },$ while $\mu >0$ is a
dimensionless positive parameter. As before, this parameter can be
associated with different complexities of the post-quarantine dynamics.
Under the condition (\ref{C2}), taking $n(t_{q})\gg 1,$ we get $\lambda =%
\tilde{c}t_{q}^{\nu -\mu }\exp [\tilde{\gamma}t_{q}^{\mu }/\mu ].$

The solution of Eq. (\ref{eneI}) can be written as 
\begin{subequations}
\label{eneISolution}
\begin{equation}
n_{I}(t)=n_{I}(t_{q})\exp \Big{[}-\frac{(t-t_{q})}{\tilde{\tau}}\Big{]}\hat{n%
}(t),\ \ \ \ \ \ \ t_{q}\leq t,
\end{equation}%
where the auxiliary function $\hat{n}(t)$ reads 
\begin{equation}
\hat{n}(t)=\exp \Big{\{}\frac{\tilde{c}_{\nu }}{\tilde{\gamma}_{\mu }}\Big{[}%
1-\exp \Big{(}-\frac{\tilde{\gamma}}{\mu }(t^{\mu }-t_{q}^{\mu })\Big{)}%
\Big{]}\Big{\}}.
\end{equation}%
For shortening the expression we defined the rates $\tilde{c}_{\nu }\equiv 
\tilde{c}t_{q}^{\nu -1},$ and $\tilde{\gamma}_{\mu }\equiv \tilde{\gamma}%
t_{q}^{\mu -1}.$ In Eq. (\ref{eneISolution}) the fitting parameters are $%
\tilde{\gamma},$ $\tilde{\tau},$ and $\mu .$ The initial condition $%
n_{I}(t_{q})=n(t_{q})$ follows from the fitting function Eq.~(\ref{Sol}).

The functional form of $n_{I}(t)$ always assumes an extremum maximal value,
whose time of occurrence is denoted as $t_{M},$ $%
(d/dt)n_{I}(t)|_{t=t_{M}}=0. $ This time can be obtained in an analytical
way in the case $\mu =1.$ From Eq.~(\ref{eneISolution}) we get 
\end{subequations}
\begin{equation}
t_{M}=t_{q}+\frac{1}{\tilde{\gamma}}\ln (\tilde{c}_{\nu }\tilde{\tau}),\ \ \
\ \ \ \mu =1,  \label{MaximalEpidemic}
\end{equation}%
which leads to $n_{I}(t_{M})=n_{I}(t_{q})[e^{(\tilde{c}_{\nu }\tilde{\tau}%
-1)}/\tilde{c}_{\nu }\tilde{\tau}]^{1/\tilde{\gamma}\tilde{\tau}}.$ For $\mu
\neq 1,$ a trascendental equation determines $t_{M}.$

The solution of Eq. (\ref{eneERE}) can be written as%
\begin{equation}
n_{R}(t)=n_{R}(t_{q})+\frac{1}{\tilde{\tau}}\int_{t_{q}}^{t}dt^{\prime
}n_{I}(t^{\prime }),\ \ \ \ \ \ \ t_{q}\leq t,  \label{EneRIntegral}
\end{equation}%
where the initial condition is $n_{R}(t_{q})=0.$ In general, the integral
can be performed in a numerical way \cite{numerico}. Complementarily, a
series solution can be obtained for $\mu =1.$ From Eq.~(\ref{eneISolution})
we get\ $(t_{q}\leq t)$%
\begin{eqnarray}
n_{R}(t) &=&n_{R}(t_{q})+n_{I}(t_{q})e^{\tilde{c}_{\nu }/\tilde{\gamma}%
}\sum_{n=0}^{\infty }\frac{1}{n!}\Big{(}\frac{-\tilde{c}_{\nu }}{\tilde{%
\gamma}}\Big{)}^{n}\frac{1}{1+n\tilde{\gamma}\tilde{\tau}}  \notag \\
&&\times \Big{\{}1-\exp \Big{[}-(1+n\tilde{\gamma}\tilde{\tau})\frac{%
(t-t_{q})}{\tilde{\tau}}\Big{]}\Big{\}}.  \label{EneRSolution}
\end{eqnarray}

The total number of infected $n(t)$ can be determined from Eq.~(\ref%
{Conservation}) and the solutions defined by Eqs.~(\ref{eneISolution}) and (%
\ref{EneRIntegral}). Its saturation value is $\lim_{t\rightarrow \infty
}n(t)=\lim_{t\rightarrow \infty }n_{R}(t).$ For $\mu =1,$ from Eq. (\ref%
{EneRSolution}) it follows%
\begin{equation}
\lim_{t\rightarrow \infty }n(t)=n(t_{q})e^{\tilde{c}_{\nu }/\tilde{\gamma}}%
\Big{(}1-\frac{\tilde{c}_{\nu }/\tilde{\gamma}}{1+\tilde{\gamma}\tilde{\tau}}%
+\cdots \Big{)},\ \ \ \ \mu =1.  \label{EneSaturation}
\end{equation}%
Consistently, the saturation value depends on both the pre- and
post-quarantine parameters.

\section{Data analysis for COVID-19}

In this section, on the basis of the proposed evolution, we analyze the
(2020) COVID-19 pandemic in different countries. For times before and after
quarantine the predicted behaviors are given by Eqs. (\ref{Sol}) and (\ref%
{Conservation}) respectively. The number of infected and dead in each day
are taken from "\textit{Our World in Data"} \cite{OWD}, while the number of
recovered are taken from \cite{DM}, which are based on official data
provided by each country. Furthermore, the (approximate) times of the
quarantine onset $t_{q}$\ are taken from public information. As a result of
standard fitting techniques, in a first step, the data previous to the
quarantine implementation allows us to obtain the parameters $\tilde{c},$ $%
\tilde{a},$ and $\nu .$ In a second step, the parameters $\tilde{\gamma},$ $%
\tilde{\tau},$ and $\mu $\ are chosen for fitting the data after the
quarantine imposition. The starting date of the epidemic is considered to be
12/31/19 in Wuhan-China, after which it becomes a pandemic spreading around
the entire world. The data of disease reported in this work is until
04/11/2020.

We consider the European countries Italy, Spain, France, United Kingdom
(UK), and Argentina. In these cases the quarantine has been applied strictly
as a unified national policy. Among these countries, Italy and Spain
reported collapse of the national health systems, which may be related with
a delay in the implementation of the quarantine. 
\begin{table}[th]
\centering
\begin{tabular}{|c|c|c|c|c|c|}
\hline
Country & Start of & $t_{q}$ & $\tilde{c}$ & $\tilde{a}$ & $\nu$ \\ 
& infections & $day$ & 1/$(day)^{\nu}$ & 1/$(day)^{\nu}$ &  \\ \hline
Italy & 02/20 & 25 & 0.7540 & 9.2588 & 0.5 \\ \hline
Spain & 02/23 & 27 & 1.0737 & 0.3160 & 0.5 \\ \hline
France & 02/26 & 25 & 0.8246 & 2.8634 & 0.5 \\ \hline
UK & 02/27 & 25 & 0.4033 & 3.4838 & 0.78 \\ \hline
Argentina & 03/04 & 22 & 0.1986 & 0.7097 & 1 \\ \hline
\end{tabular}%
\caption{Fit parameters obtained by using Eq. (\protect\ref{Sol}) for
European countries and Argentina in the pre-quarantine period. Dates
correspond to this year (2020).}
\label{PreParametros}
\end{table}
%
\begin{table}[h]
\centering
\begin{tabular}{|c|c|c|c|c|}
\hline
Country & $\tilde{\gamma}$ & $1/\tilde{\tau}$ & $\mu $ & $f $ \\ 
& 1/$(day)^{\mu}$ & 1/$(day)^{\mu}$ &  &  \\ \hline
Italy & 0.050 & 0.027 & 1 & 0.70 \\ \hline
Spain & 0.053 & 0.055 & 1 & 0.74 \\ \hline
France & 0.030 & 0.055 & 1 & 0.42 \\ \hline
UK & 0.030 & 0.020 & 1 & 0.42 \\ \hline
Argentina & 0.055 & 0.032 & 1 & 0.77 \\ \hline
\end{tabular}%
\caption{Fit parameters obtained by using Eq. (\protect\ref{Conservation})
jointly with the solutions (\protect\ref{eneISolution}) and (\protect\ref%
{EneRSolution}), for European countries and Argentina in the post-quarantine
period. The quarantine parameter $f$ is defined by Eq.~(\protect\ref{Efe}).}
\label{PostParametros}
\end{table}
%
In Table \ref{PreParametros} the epidemic starting date (month/day) in each
country is provided. It was taken as the date where an effective increase is
observed in the reported data. Furthermore, the quarantine times $t_{q}$
were obtained from public information. We also include the results of the
fitting parameters $\{\tilde{c},$ $\tilde{a},$ $\nu \}$ associated to Eq. (%
\ref{Sol}). Table \ref{PostParametros} summarizes the fitting parameters $\{%
\tilde{\gamma},$ $\tilde{\tau},$ $\mu \}$ associated to Eq. (\ref%
{Conservation}), with solutions (\ref{eneISolution}) and (\ref{EneRIntegral}%
). 
\begin{figure}[tb]
\center
\includegraphics[bb=26 70 725 500,angle=0,width=8.5cm]{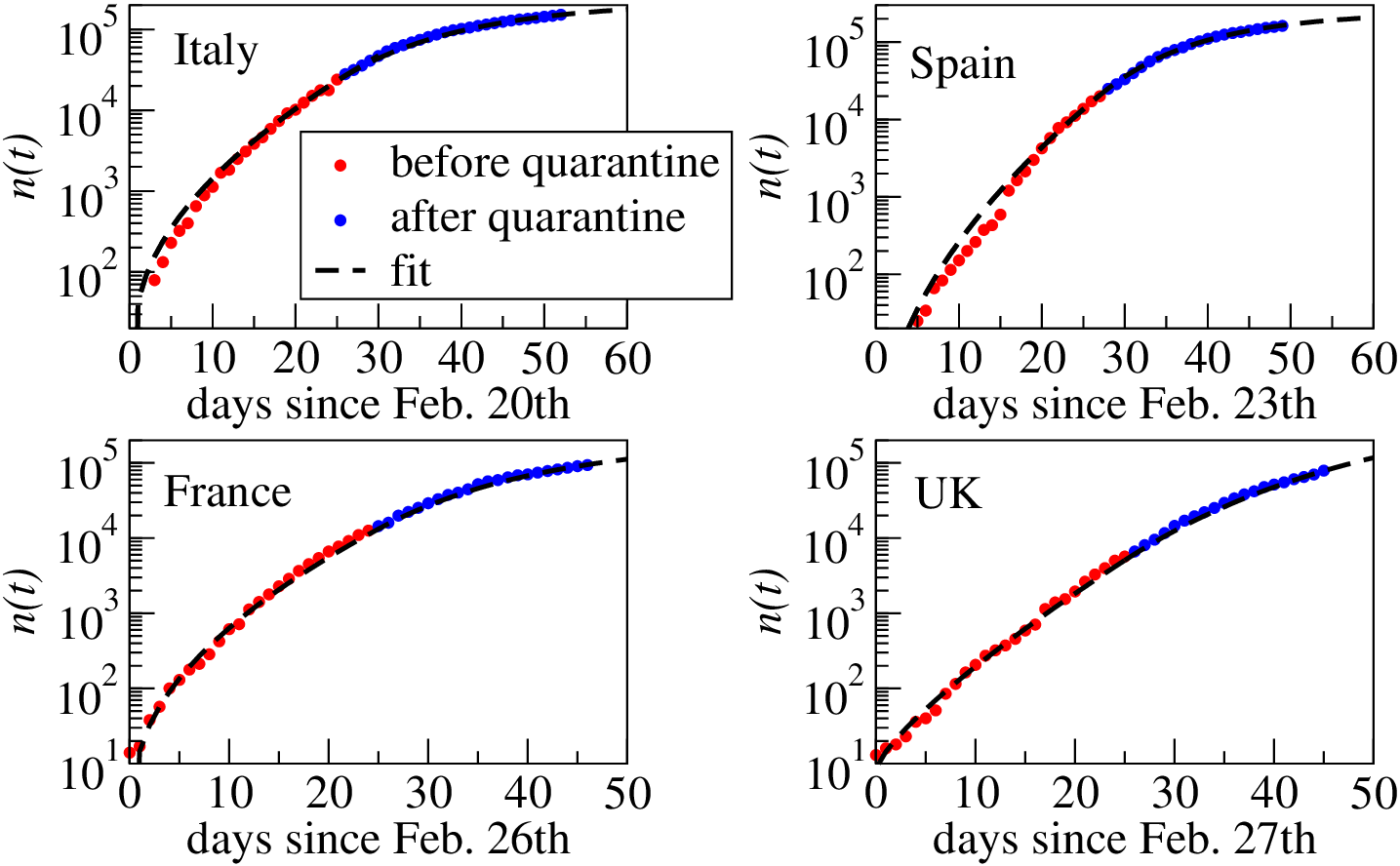}
\caption{Evolution of the total number $n(t)$ of COVID-19 infections
reported for Italy, Spain, France, and UK. The symbols are the data while
the dashed lines are the analytical fits based on Eqs.~(\protect\ref{Sol})
and (\protect\ref{Conservation}). The red and blue colors indicates times
before and after the quarantine respectively.}
\label{EuropaLog}
\end{figure}

For the European countries, in Fig. \ref{EuropaLog} it is shown the
(available) data and the performed fits for the total number of infected $%
n(t).$ Even in a logarithmic scale, the model provides a very good fitting
of the reported data. From Table \ref{PreParametros}, it follows that the
dynamics of virus propagation in the first pre-quarantine stage [Eq. (\ref%
{Sol})] is not exponential, $\nu \neq 1.$ For Italy, Spain and France, the
growth is proportional to $t^{1/2},$ while for UK is proportional to $%
t^{0.78}.$ This last value is consistent with a very pronounced increase of
the number of infections at the end of the data. On the other hand, in the
post quarantine period we find that the best fitting of the available data
is with $\mu \approx 1$ [Table \ref{PostParametros}], corresponding to the
solutions (\ref{eneISolution}) and (\ref{EneRSolution}). This feature is
valid for all countries, which shows that the epidemic after quarantine
assumes the same dynamics.

The previous results provide a strong support to our modeling, which in a
first stage is able to capture departures with respect to a pure exponential
growth, as well as a posterior development of a universal dynamics. 
\begin{figure}[tb]
\center
\includegraphics[bb=45 55 710 510,angle=0,width=8.5cm]{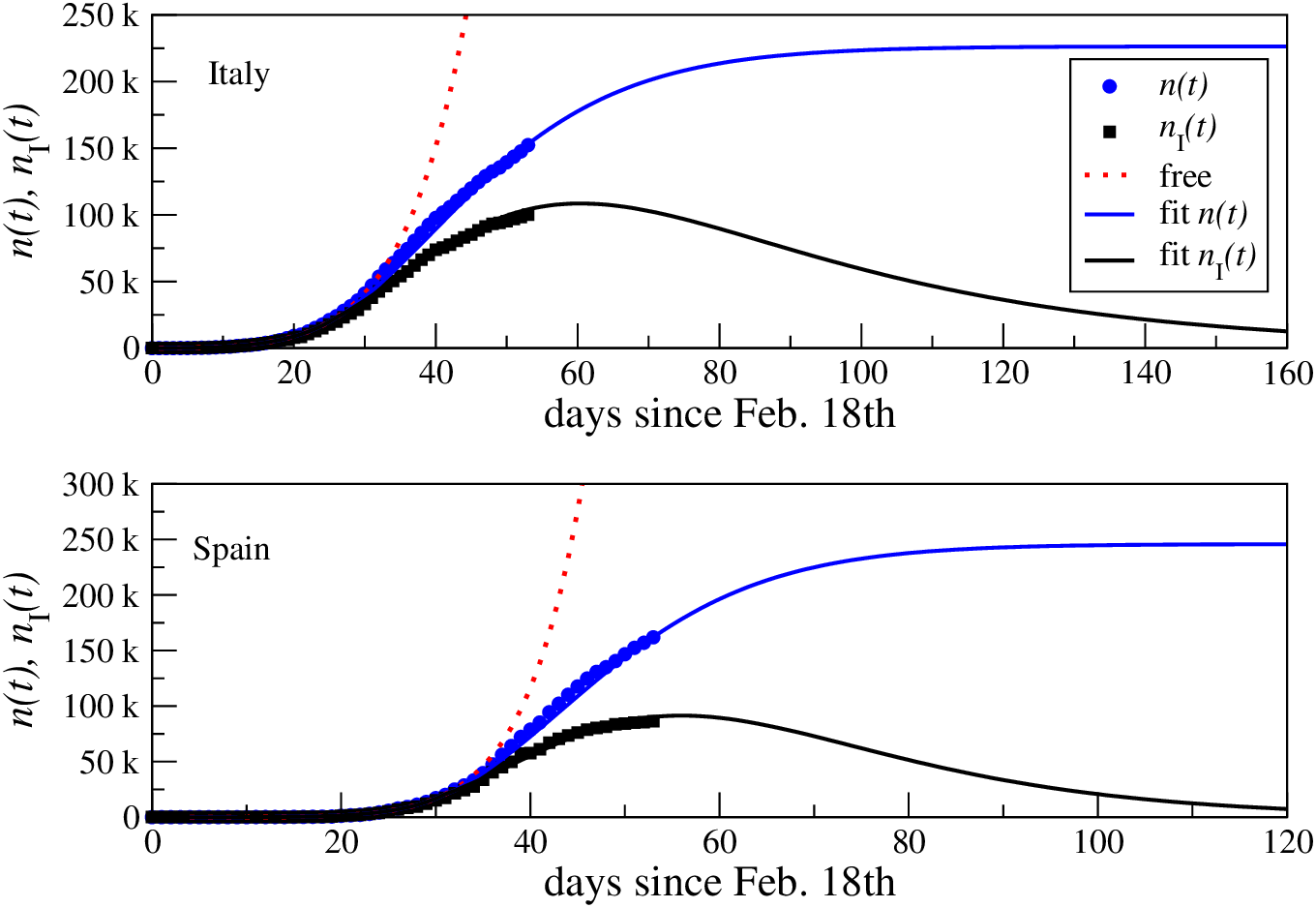} %
\includegraphics[bb=45 55 710 610,angle=0,width=8.5cm]{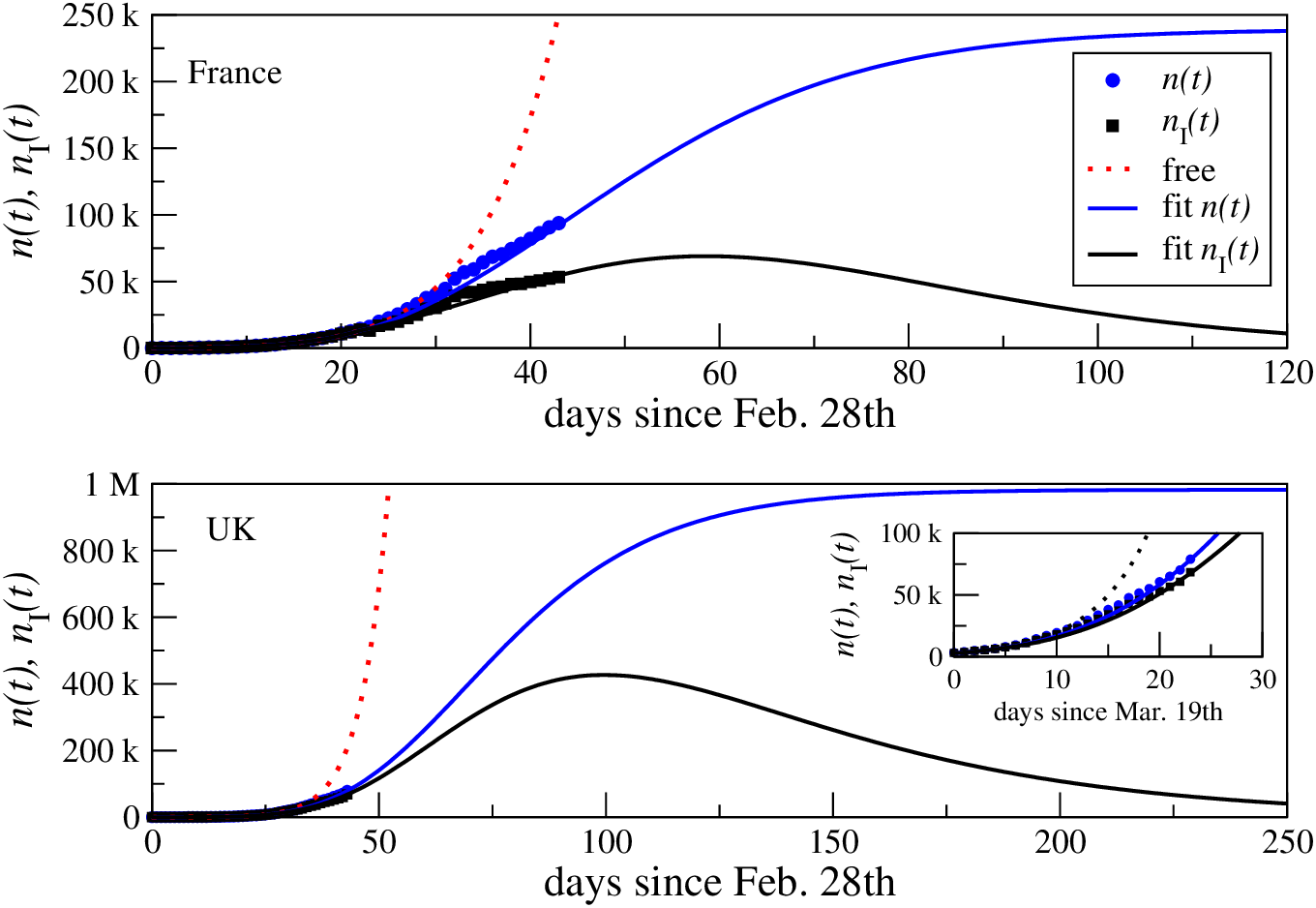}
\caption{Estimated temporal evolution of $n(t)$ [Eqs.~(\protect\ref{Sol})
and (\protect\ref{Conservation})] and $n_{I}(t)$ [Eqs.~(\protect\ref%
{eneISolution}) and (\protect\ref{EneRSolution})] for European countries.
The symbols are the data while the full lines are the analytical fits. The
dotted lines correspond to the (free) unconstrained behavior without
quarantine.}
\label{ItaliaEspania}
\end{figure}

In a linear time-scale, in Fig. \ref{ItaliaEspania} we show the predicted
time evolution of $n(t)$ and $n_{I}(t)$ jointly with the available data. It
can clearly be seen that the quarantine turns out to be an effective policy
that reduces infections considerably. In fact, with quarantine (full line) $%
n_{I}(t)$ develops a maximal value and departs considerably from the
unconstrained (free) evolution (dotted line) Eq. (\ref{Sol}). For Italy,
Spain and France, the peak of the epidemic is estimated to develop in 55-60
days after the epidemic beginning. Instead, for the UK case, the peak
emerges after 90 days. These estimations follow from Eq. (\ref%
{MaximalEpidemic}). On the other hand, in all cases, the saturation value of 
$n(t)$ can be very well approximated by the first term of Eq.$~$(\ref%
{EneSaturation}).

\begin{figure}[tb]
\center
\includegraphics[bb=25 175 715 510,angle=0,width=8.5cm]{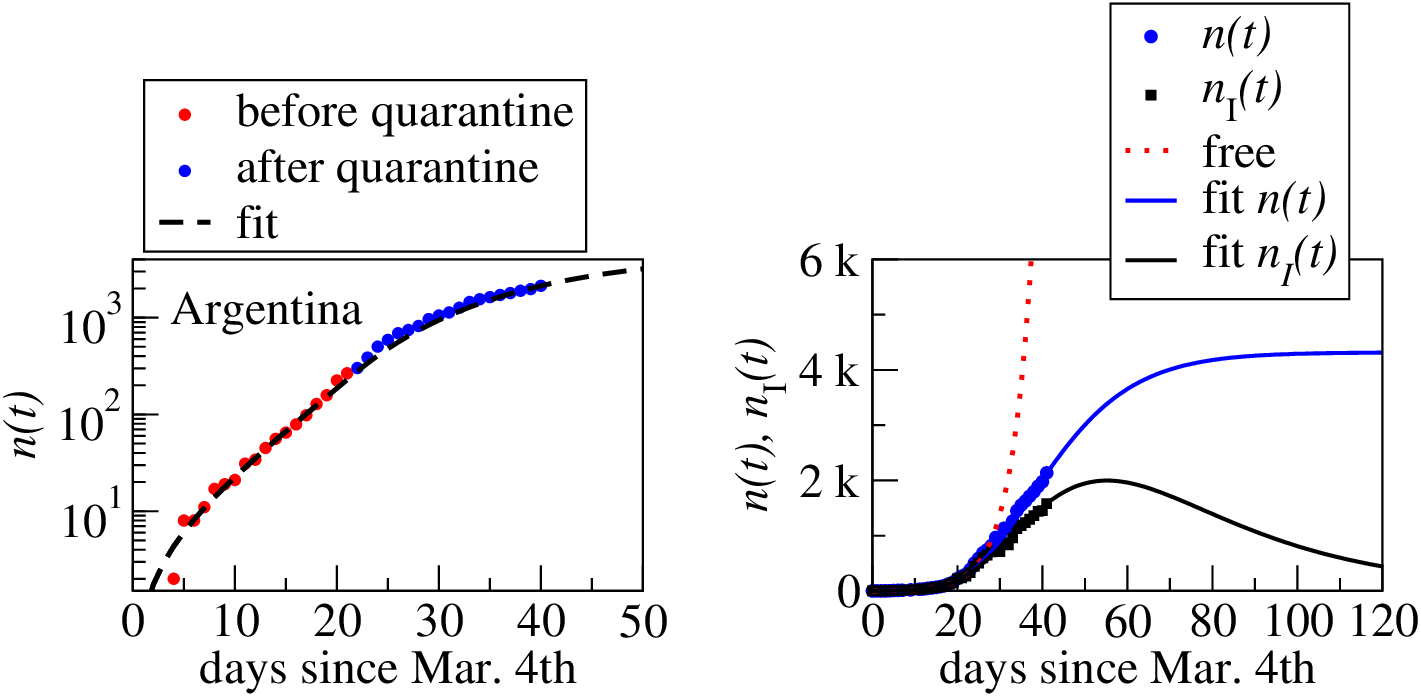}
\caption{Evolution of the number of COVID-19 infections for Argentina. (a)
Reported data (symbols). Full line is the analytical fit. The red and blue
colors indicates times before and after the quarantine respectively. (b)
Predicted behavior of $n(t)$ and $n_{I}(t)$ in a longer time scale (full
line), jointly with unconstrained (free) one (dotted line).}
\label{Argentina}
\end{figure}

In Fig. \ref{Argentina} we show the Argentinean case, country that also
follows a severe quarantine. Contrarily to the European cases, here the
epidemic growth before the quarantine is exponential $(\nu =1,$ Table \ref%
{PreParametros}). During the quarantine, a considerable decrease in the
growing velocity is observed. Given that this country implemented the
quarantine when the number of reported cases hadn't increase considerably, a
small level of infection is predicted at the epidemic peak when compared
with the previous countries. The maximal number of infected cases can also
be very well approximated by the first term of Eq.~(\ref{EneSaturation}).

With the information of Tables \ref{PreParametros} and \ref{PostParametros},
it is possible to estimate an important parameter that characterizes the
propagation dynamics of the disease, this is the so-called reproduction
number $R(t)\equiv n(t+\alpha )/n(t)$ [Eq. (\ref{Sol})]. The time interval $%
\alpha $ is the number of days for which $R(t)$ acquires a certain value. In
the case of Argentina, in the pre-quarantine period, taking for example $%
R(t)=2$ (doubling the number of infections) it is possible to obtain $\alpha
\approx {\tilde{c}^{-1}}ln(2)=3.5$ days, which is in accordance with the
values of 3-4 days obtained from the infections reported at the beginning of
the spread of the disease. Making a similar analysis in the post-quarantine
stage, the new value of $\alpha $ for which $R(t)=2$ is given by $\alpha
\approx {\tilde{\gamma}^{-1}ln(2)}=12.6$ days. This result is also in
agreement with the values of 10-11 days obtained from reported data. The
change in the value of $R(t)$ $(3.5\rightarrow 12.6$ days), clearly reflects
the beneficial effect of the quarantine in decreasing the spread of the
epidemic. For European countries a similar analysis can be done. 

\section{Quarantine relaxation scenarios}

The quarantine policy is one of the most appropriate mechanisms to control
an epidemic in a social way. Nevertheless, this kind of solution cannot be
maintained during large periods of time. Therefore, establishing simple
criteria that allow softening tightness of the quarantine without implying a
collapse of the health system, is a problem whose characterization is
demanded.

In a roughly way, each quarantine scenario can be defined by the fraction $%
f\in (0,1)$ of the total population that is completely isolated. Thus, a
quarantine softening can be related to a change in this parameter. In our
minimal model, the parameters associated to the quarantine period are the
time-dependent rate $r_{q}(t)$ and the time $\tilde{\tau}$ [see Eqs. (\ref%
{eneI}) and (\ref{eneERE})]. Given that $\tilde{\tau}$ defines the time
scale for the transition $n_{I}(t)\rightarrow n_{R}(t),$ we consider that it
has a weak dependence on $f.$ Nevertheless, it has a strong regional
dependence [see Table \ref{PostParametros}].

For the analyzed countries $r_{q}(t)$ is defined by Eq. (\ref{EreQ}) with $%
\mu =1.$ Therefore, the unique parameter that can be related with $f$ is the
rate parameter $\tilde{\gamma},$ which defines the time scale of the decay
of $r_{q}(t).$ As an ansatz, we relate both parameters as follows%
\begin{equation}
\frac{1}{\tilde{\gamma}}=\frac{t_{I}}{f}.  \label{Efe}
\end{equation}%
Here, $t_{I}$ is the average number of days it takes a person to develop
health signals of the disease. Consistently a higher $f$ leads to a higher $%
\tilde{\gamma},$ which from Eq. (\ref{EneSaturation}) implies a lowering in $%
\lim {}_{t\rightarrow \infty }n(t).$ In Table \ref{PostParametros} we
summarize the values of $f$ that follow from Eq. (\ref{Efe}) taking $%
t_{I}=14 $ days \cite{expert}. These values give a roughly estimation of the
quarantine tightness implemented by each country.

The relation (\ref{Efe}) allows us to analyze in a qualitative way the
changes in the epidemic dynamics under a governmental softening of the
quarantine policy. Each scenario corresponds (or can be related) to a
different value of $f\rightarrow f_{p}<1,$ which now measures departures (or
relaxation) from a more strict quarantine policy, $f_{p}<f.$ The inverse
case $f_{p}>f$ corresponds to an increasing of the quarantine tightness. The
problem is to estimate the new epidemic peak (if it develops) and new
epidemic saturation value as a function of $f_{p}$ and the time $t_{p}$ at
which the quarantine\ relaxation begins $(t_{p}>t_{q}).$

We assume that for times $t\geq t_{p}$ the evolution remains the same [Eqs.~(%
\ref{eneI}) and (\ref{eneERE})], being defined with updated parameters.
Thus, the expressions for $n_{I}(t)$ and $n_{R}(t)$ are given by Eqs. (\ref%
{eneISolution}) $(\mu =1)$ and (\ref{EneRSolution}) respectively, under the
replacements%
\begin{equation}
t_{q}\rightarrow t_{p},\ \ \ \ \ \ \ \ \ \ \ \ \tilde{c}_{\nu }\rightarrow 
\tilde{c}_{\nu }e^{-\tilde{\gamma}(t_{p}-t_{q})}.  \label{Update1}
\end{equation}%
The replacement $\tilde{c}_{\nu }\rightarrow \tilde{c}_{\nu }e^{-\tilde{%
\gamma}(t_{p}-t_{q})}$ guarantees that the evolution remains the same when
the translation $t_{q}\rightarrow t_{p}$ is introduced. In a second step, in
all places the rate parameter $\tilde{\gamma}$ is updated as follows%
\begin{equation}
\tilde{\gamma}\rightarrow \tilde{\gamma}_{p}\equiv \frac{f_{p}}{t_{I}},
\label{Update2}
\end{equation}%
which introduces the change in the quarantine policy. It is simple to check
that when $f_{p}=f,$ the epidemic dynamics remains unaltered or invariant.

By introducing the replacements (\ref{Update1}) and (\ref{Update2}) into
Eq.~(\ref{MaximalEpidemic}), after some algebra it is simple to obtain the
time $t_{M_{p}}$ at which $n_{I}(t)$ may develop (or not) a second peak,%
\begin{equation}
t_{M_{p}}=t_{q}+\frac{1}{\tilde{\gamma}_{p}}\ln (\tilde{c}_{\nu }\tilde{\tau}%
).  \label{NewTM}
\end{equation}%
In fact, considering that the condition $t_{p}<t_{M_{p}}$ must be fulfilled,
a second peak emerges if the following inequality is satisfied%
\begin{equation}
\frac{t_{p}-t_{q}}{t_{I}}<\frac{1}{f_{p}}\ln (\tilde{c}_{\nu }\tilde{\tau}).
\label{Condition}
\end{equation}%
Thus, when this inequality is satisfied a second epidemic peak is observed
in the interval $t>t_{p}.$ When the condition (\ref{Condition}) is not
satisfied the epidemic does not develop a second peak in the interval $%
t>t_{p}.$ 
\begin{figure}[tbh]
\center
\includegraphics[bb=40 70 740 510,angle=0,width=8.5cm]{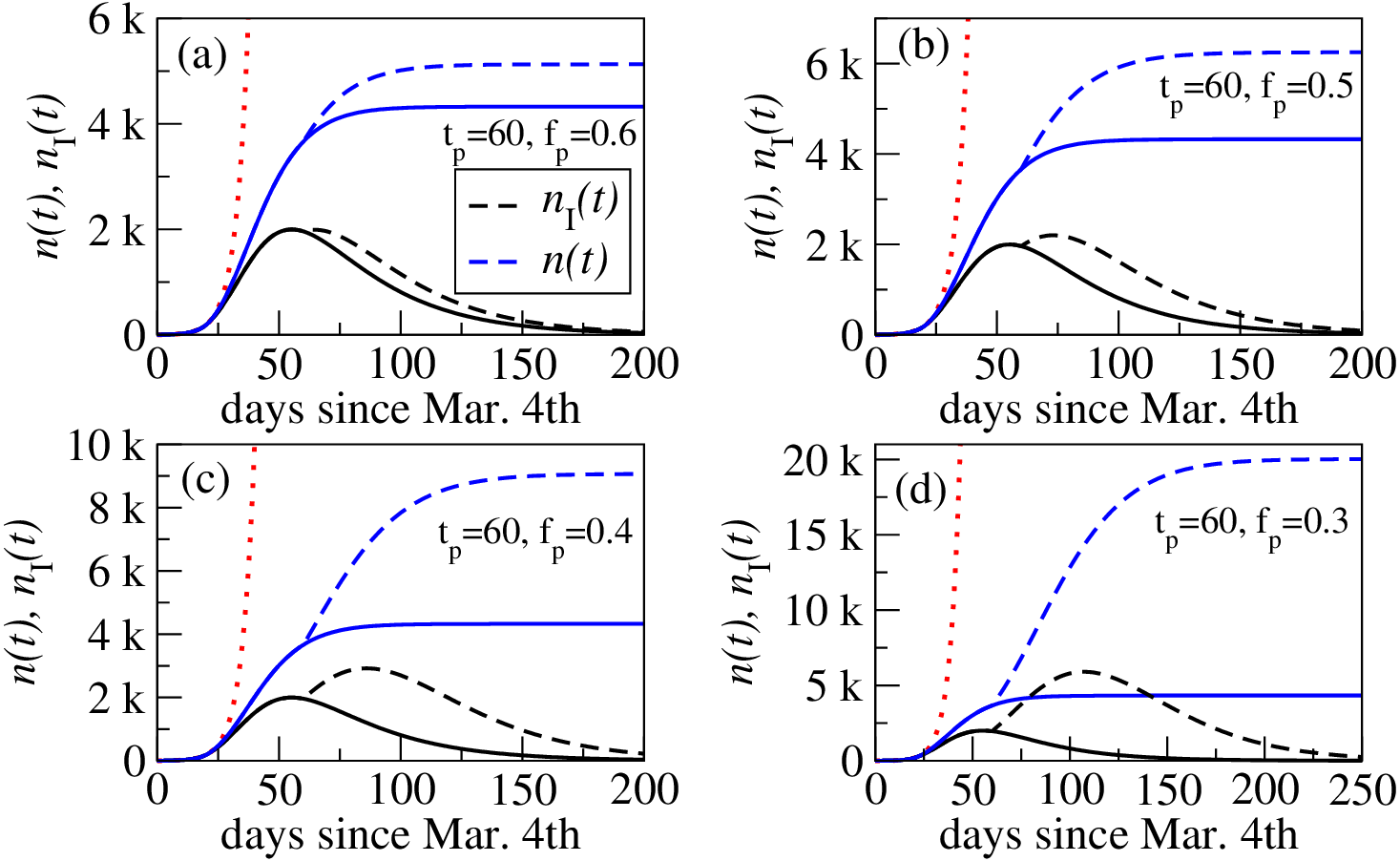}
\caption{Estimated temporal evolution of $n(t)$ and $n_{I}(t)$ for different
scenarios with quarantine softening (see text) for the Argentinean case. The
solid curves correspond to the original dynamics (Fig. \protect\ref%
{Argentina}) where $f=0.77.$ In (a) to (d) it is taken $f_{p}=0.6,$ $0.5,$ $%
0.4,$ and $0.3$ respectively. In all cases $t_{p}=60$ days. The dotted lines
correspond to the (free) behavior without quarantine.}
\label{PostCO19}
\end{figure}

After quarantine softening, the saturation value of \ the total number of
infected follows again from the relation $\lim_{t\rightarrow \infty
}n(t)=\lim_{t\rightarrow \infty }n_{R}(t).$ From Eq. (\ref{EneRSolution}),
by using the update rules (\ref{Update1}) and (\ref{Update2}) we get%
\begin{eqnarray}
\lim_{t\rightarrow \infty }n(t) &=&n_{R}(t_{p})+n_{I}(t_{p})\exp [\tilde{c}%
_{\nu }\tilde{\gamma}_{p}^{-1}e^{-\tilde{\gamma}_{p}(t_{p}-t_{q})}]  \notag
\\
&&\times \Big{(}1-\frac{\tilde{c}_{\nu }\tilde{\gamma}_{p}^{-1}e^{-\tilde{%
\gamma}_{p}(t_{p}-t_{q})}}{1+\tilde{\gamma}_{p}\tilde{\tau}}+\cdots \Big{)}.
\label{MaximalPost}
\end{eqnarray}

In order to enlighten the previous results, in Fig. (\ref{PostCO19}) we
analyze different possible post-quarantine scenarios for the Argentinean
case [Fig. \ref{Argentina}]. The characteristic parameters before and after
the quarantine are those of Table \ref{PreParametros} and \ref%
{PostParametros} respectively. In particular, $f=0.77.$

In Figs. \ref{PostCO19}(a) to (d) we take a fixed $t_{p}=60$ days while $%
f_{p}=0.6,$ $0.5,$ $0.4,$ and $0.3.$ As expected, in all cases, under the
quarantine softening $f\rightarrow f_{p}$ the maximal number $%
[\lim_{t\rightarrow \infty }n(t)]$ of total infected increases [Eq.~(\ref%
{MaximalPost})]. This change starts to be significant for smaller values of $%
f_{p},$ where most of the population recovers its mobility. In addition, a
diminishing in $f_{p}$ also leads to the appearance of extra epidemic peaks
in the number of active infected $n_{I}(t)$, whose time of occurrence can be
inferred from Eq.~(\ref{NewTM}). Due to the condition (\ref{Condition}), for
the higher value of $f_{p}$ the extra peak does not develop.

\section{Summary and Conclusions}

Given the international network of public transportation, the epidemic
growth in a given city or country may be triggered by external agents that
arrive from an infected region. This first stage is not universal. After
that, local governmental policies such as quarantine ones are taken in order
to mitigate the epidemic growth. We presented a minimal model were all these
components are taken into account, being defined by parameters that can be
deduced from public information and fitting of the available reported data.

The first stage, that is the importation of external cases plus the
beginning of local contagions, was modeled by a linear dynamics [Eq. (\ref%
{ModelShort})]. The proposed solution [Eq.~(\ref{Sol})] is able to fit
different sub, super, as well as standard exponential behaviors. The
quarantine period was modelled by splitting the total number of infected in
active and inactive (recovered and dead) [Eq.~(\ref{Conservation})], which
dynamics can be read as a linearization of a SIR like model [Eqs. (\ref{eneI}%
) and (\ref{eneERE})].

The model was applied to the COVID-19 pandemic, analyzing reported data in
different countries. For European countries such as Italy, Spain, France,
and UK, the model provides a very good fitting of the available data. In
addition, estimations for the corresponding epidemic peaks were obtained.
Similarly to the case of Argentina, these countries established quarantine
as a rigorous national policy. While in the pre-quarantine period a
universal behavior is not observed, this is the case in the post-quarantine
stage, where independently of the country the epidemic can be fitted with
the same complexity parameter. These features support the splitting
introduced in our model, as well as the estimations obtained from it.

The present approach also allows to analyze possible quarantine softening
scenarios. The dynamics remains the same, being defined with a modified rate
that depends on a factor that is proportional to the fraction of the total
population under quarantine. The formalism furnishes roughly estimations for
the changes in the epidemic peak and saturation values. Thus, we conclude
that the present contribution may provide a valuable tool for taking
relevant decisions about epidemic control. In particular, it allows to
obtain qualitative estimations of the degree of quarantine softening in
order to avoid a posterior collapse of a given health system.

\section*{Acknowledgments}

The authors thank Orlando Billoni for sending relevant references and
information. Also to Mariano Bonifacio and Esteban Sorocinschi for a
critical reading of the manuscript. L.A.R.P. and A.A.B. thanks support from
CONICET, Argentina.

\appendix*

\section{Relation with non-linear SIR like-models}

In the recent literature there are alternative generalizations of SIR models
that take into account different governmental quarantine policies \cite%
{berlin,niteroi,Schaback,Ferguson,Zhang,Read,wang,ritter,ramos,yang,lopez,peng,garcia}%
. For example, from \cite{berlin}, we write 
\begin{subequations}
\label{SIR}
\begin{eqnarray}
\frac{dS}{dt} &=&-\alpha SI-\kappa _{0}S, \\
\frac{dI}{dt} &=&\alpha SI-\beta I-\kappa _{0}I-\kappa I.
\end{eqnarray}%
$S$ and $I$ are the number of susceptible and infected persons respectively.
The rate $\kappa _{0}$ measures the removed agents due to quarantine, while
the rate $\kappa $ measures the removed infected people. The constant $%
\alpha $ and $\beta $ have the usual meaning in SIR models \cite%
{murray,networks}.

The previous non-linear evolution can be linearized under the condition $%
\alpha SI\ll \kappa _{0}S.$ Thus, it follows that 
\end{subequations}
\begin{equation}
S\simeq s_{0}\exp (-\kappa _{0}t).  \label{Aproxi}
\end{equation}%
Introducing Eq. (\ref{Aproxi}) into Eq. (\ref{SIR}) we get%
\begin{equation}
\frac{dI}{dt}\simeq \lbrack \alpha s_{0}e^{-\kappa _{0}t}-(\beta +\kappa
_{0}+\kappa )]I.
\end{equation}%
Under the mapping $r_{q}(t)=\alpha s_{0}e^{-\kappa _{0}t}$ and $\tilde{\tau}%
^{-1}=(\beta +\kappa _{0}+\kappa ),$ we recover the proposal Eq. (\ref{eneI}%
), $n_{I}(t)\leftrightarrow I.$


\begin{thebibliography}{99}
\bibitem{murray} J. D. Murray, \textit{Mathematical Biology, An Introduction}%
, Vol. I and II, Springer (2001).

\bibitem{networks} M. E. J. Newman,\textit{\ Networks, and Introduction},
Oxford (2010).

\bibitem{berlin} B. F. Maier and D. Brockmann, Effective containment
explains sub-exponential growth in confirmed cases of recent COVID-19
outbreak in Mainland China, medRxiv preprint doi:
https://doi.org/10.1101/2020.02.18.20024414.

\bibitem{niteroi} N. Crokidakis, Data analysis and modeling of the evolution
of COVID-19 in Brazil, arXiv:2003.12150v1 [q-bio.PE] 26 Mar 2020.

\bibitem{Schaback} R. Schaback, Modelling Recovered Cases and Death
Probabilities for the COVID-19 Outbreak, arXiv:2003.12068v1 [q-bio.PE] 26
Mar 2020.

\bibitem{Ferguson} N. M. Ferguson, \textit{et. al.},\ Imperial College
COVID-19 Response Team, Impact of non-pharmaceutical interventions (NPIs) to
reduce COVID-19 mortality and healthcare demand, DOI:
https://doi.org/10.25561/77482.

\bibitem{Zhang} J. Zhang \textit{et. al.}, Age profile of susceptibility,
mixing, and social distancing shape the dynamics of the novel coronavirus
disease 2019 outbreak in China, medRxiv preprint doi:
https://doi.org/10.1101/2020.03.19.20039107.

\bibitem{Read} J. M. Read, J. R. E. Bridgen, D. A.T. Cummings, A. Ho, C. P.
Jewell, Novel coronavirus 2019-nCoV: early estimation of epidemiological
parameters and epidemic predictions, medRxiv preprint doi:
https://doi.org/10.1101/2020.01.23.20018549.

\bibitem{wang} C. Wang \textit{et. al.}, Evolving Epidemiology and Impact of
Non-pharmaceutical Interventions on the Outbreak of Coronavirus Disease 2019
in Wuhan, China, medRxiv preprint doi:
https://doi.org/10.1101/2020.03.03.20030593.

\bibitem{ritter} M. Ritter, J. -D. Haynes, and K. Ritter, Covid-19 -- A
simple statistical model for predicting ICU load in exponential phases of
the disease.

\bibitem{ramos} B. Ivorra, M.R. Ferr\'{a}ndez, M. Vela-P\'{e}rez, and A.M.
Ramos. Mathematical modeling of the spread of the coronavirus disease 2019
(COVID-19) taking into account the undetected infections. The case of China,
Research Gate (2020).

\bibitem{yang} Ruiyun Li, Sen Pei, Bin Chen, Yimeng Song, Tao Zhang,Wan
Yang, and Jeffrey Shaman, Substantial undocumented infection facilitates the
rapid dissemination of novel coronavirus (sars-cov2), Science (2020).

\bibitem{lopez} Leonardo L\'{o}pez and Xavier Rod\'{o}, A modified SEIR
model to predict the COVID-19 outbreak in Spain: Simulating control
scenarios and multi-scale epidemics, medRxiv (2020).

\bibitem{peng} Liangrong Peng, Wuyue Yang, Dongyan Zhang, Changjing Zhuge,
and Liu Hong, Epidemic analysis of COVID-19 in China by dynamical modeling,
(2020).

\bibitem{garcia} V\'{\i}ctor M. P\'{e}rez-Garc\'{\i}a. Relaxing quarantine
after an epidemic: A mathematical study of the spanish covid-19 case, 2020.
DOI: 10.13140/RG.2.2.36674.73929/1.

\bibitem{numerico} For the analyzed data, we performed this calculus by
using different integration algorithms.

\bibitem{OWD} Our World in Data: https://ourworldindata.org/corona
virus-source-data

\bibitem{DM} https://datosmacro.expansion.com/otros/coronavirus

\bibitem{expert} https://www.medrxiv.org/content/10.1101/2020.03.15.20
036533v1.full.pdf+html
\end{thebibliography}
\end{document}